\documentclass[lettersize,journal]{IEEEtran}
\usepackage{subcaption}
\usepackage{amsmath,amsfonts}
\usepackage{algorithmic}
\usepackage{algorithm}
\usepackage{array}
\usepackage[caption=false,font=normalsize,labelfont=sf,textfont=sf]{subfig}
\usepackage{textcomp}
\usepackage{stfloats}
\usepackage{url}
\usepackage{verbatim}
\usepackage{graphicx}
\usepackage{cite}
\begin{document}

\title{Comprehensive Evaluation of Quantitative Measurements from Automated Deep Segmentations of PSMA PET/CT Images}

\author{
    Obed Korshie Dzikunu\textsuperscript{1,2,3}*,  
    Amirhossein Toosi\textsuperscript{1,2}\dag,  
    Shadab Ahamed\textsuperscript{1,2}\dag,  
    Sara Harsini\textsuperscript{2},
    Fran\c{c}ois B\'{e}nard\textsuperscript{1,2},
    Xiaoxiao Li\textsuperscript{1,3},  
    Arman Rahmim\textsuperscript{1,2}\vspace{3mm}

\thanks{This work involved human subjects in its research. Approval of all ethical and experimental procedures and protocols was granted by the UBC BC Cancer Research Ethics Board under Application No. H16-01551 and performed in line with the Declaration of Helsinki.}
\thanks{\textsuperscript{1}University of British Columbia, Vancouver, Canada.}
\thanks{\textsuperscript{2}BC Cancer Research Institute, Vancouver, Canada.}
\thanks{\textsuperscript{3}Vector Institute, Toronto, Canada.}
\thanks{* Corresponding author: Obed Korshie Dzikunu (okdzikunu@gmail.com).}
\thanks{\dag These authors contributed equally.}
}
\maketitle

\begin{abstract}
This study performs a comprehensive evaluation of quantitative measurements as extracted from automated deep-learning-based segmentation methods, beyond traditional Dice Similarity Coefficient assessments, focusing on six quantitative metrics, namely SUV\(_{\text{max}}\), SUV\(_{\text{mean}}\), total lesion activity (TLA), tumor volume (TMTV), lesion count, and lesion spread. We analyzed 380 prostate-specific membrane antigen (PSMA) targeted [$^{18}$F]DCFPyL PET/CT scans of patients with biochemical recurrence of prostate cancer, training deep neural networks, U-Net, Attention U-Net and SegResNet with four loss functions: Dice Loss, Dice Cross Entropy, Dice Focal Loss, and our proposed L1 weighted Dice Focal Loss (L1DFL). Evaluations indicated that Attention U-Net paired with L1DFL achieved the strongest correlation with the ground truth (Lin's concordance correlation = 0.90–0.99 for SUV\(_{\text{max}}\) and TLA), whereas models employing the Dice Loss and the other two compound losses, particularly with SegResNet, underperformed. Equivalence testing (TOST, $\alpha$ = 0.05, $\Delta$ = $20\%$) confirmed high performance for SUV metrics, lesion count and TLA, with L1DFL yielding the best performance. By contrast, tumor volume and lesion spread exhibited greater variability. Bland-Altman, Coverage Probability, and Total Deviation Index analyses further highlighted that our proposed L1DFL minimizes variability in quantification of the ground truth clinical measures. The code is publicly available at: https://github.com/ObedDzik/pca\_segment.git.
\end{abstract}


\section{Introduction}
Prostate cancer arises from the uncontrolled proliferation of malignant epithelial cells in the prostate gland \cite{rawla2019epidemiology}. In the early stages, these cells depend on testosterone for growth, and often androgen deprivation may control the disease \cite{basaria2002long}. Patients with localized disease are often treated with curative intent using radical prostatectomy (RP) or radiotherapy (RT) \cite{van2019prognostic}. Despite these interventions, about \(20-40\%\) of patients develop biochemical recurrence (BCR), which typically involves a rise in prostate-specific antigen (PSA) \cite{suardi2008nomogram,chun2006anatomic}. However, depending on the type of primary treatment, specific definitions of BCR may be employed. For instance, after RP, BCR is indicated by two consecutive PSA values $\geq$0.2 ng/mL \cite{cookson2007variation}. Although BCR does not always signify clinical progression, it is associated with an increased risk of metastasis and mortality in certain subsets of patients \cite{simon2022best}.

In managing BCR, imaging modalities like prostate-specific membrane antigen (PSMA) positron emission tomography (PET) may be used and have substantially increased the detection of early metastatic disease \cite{afshar2015diagnostic}. Identified lesions may be manually segmented by physicians to extract various quantitative measures of clinical relevance such as total molecular tumor volume (TMTV), mean and maximum standardized uptake value (SUV\(_{\text{mean}}\) and SUV\(_{\text{max}}\)), and total lesion activity (TLA) \cite{ahamed2023comprehensive}. These metrics are increasingly being identified as important in the prediction of patient outcomes and in treatment decisions. The manual process of lesion segmentation and quantification of these metrics, however, is time-consuming, leading to an interest in artificial intelligence (AI)-based methods to automate the process.

Despite their promise, automated segmentation introduces its own unique set of challenges. Typically, lesions that resurface after treatment are small, heterogeneous in shape, location, and uptake activity, and occupy a small portion compared to the background volume. Consequently, most methods struggle to balance lesion sensitivity and specificity, often presenting high rates of false positives.

Several domain-specific techniques have thus been explored to address these challenges, including developing models for specific anatomical sites \cite{holzschuh2023deep,kostyszyn2021intraprostatic}, using SUV thresholding methods \cite{foster2014review}, and using healthy organs as anatomical priors to refine model attention \cite{li2024automated}. These methods are tailored to address specific challenges involved in the automated segmentation of prostate cancer lesions but are limited in their application. For instance, models tailored for specific anatomical sites fail to generalize well for metastatic conditions. Also, the use of SUV thresholding limits sensitivity for lesions with low avidity for the radiotracer \cite{li2024automated}. Moreover, the segmentation of healthy organs as anatomical priors requires manual quality assurance steps to ensure accurate boundary delineation \cite{li2024automated}.

More generalized domain-agnostic approaches involve the design of various loss functions which can be tailored to the segmentation task. For instance, Xu et. al (\cite{XU2023106882}) proposed a batch-wise Dice Loss function for foreground voxel weighting. While significant improvements were reported on the segmentation accuracy compared to the use of the non-weighted Dice Loss function, the overall mean Dice Similarity Coefficient (DSC) remained low at 0.43 due to high false positive rates. Typically, the Dice Loss is used for medical image segmentation tasks due to its robustness to data class imbalance, however, it is less robust in handling differences in voxel classification difficulties. That is, within the background and foreground classes, some voxels are more difficult to classify causing models to achieve suboptimal performances on these regions \cite{9338261}. Distribution-based loss functions, on the other hand, perform better at adjusting the model's focus based on voxel classification difficulty, thus are combined with region-based objective functions to form compound losses with improved robustness and generalizability \cite{zhang2021rethinking}. Yet, by combining these loss functions, a trade-off is often required since the inherent balance in sensitivity and precision preserved by the region-based loss functions tends to be lost, and precision is often traded for sensitivity \cite{YEUNG2022102026}. 

To maintain balance in recall and precision while adapting the model's focus based on the difficulty of the voxel classification, we proposed a custom loss function, L1-weighted Dice Focal loss (\cite{dzikunu2025adaptivevoxelweightedlossusing}), which adjusts focus based on the L1 norms of predicted probabilities and ground truth labels. With this loss function, higher DSC and F1 scores were achieved in detecting small-sized prostate cancer lesions than standard loss functions such as the Dice Loss and Dice Focal Loss. However, similar to other studies, whether these improvements correlate with clinical value remain an open question. To assess the potential clinical validity of these automated methods, it is vital to evaluate the ability of algorithms to yield clinically meaningful outcomes in addition to their performance on overlap metrics like DSC \cite{jha2022nuclear}.

As such, in this study, we adopt a medically relevant evaluation approach to assess the potential clinical relevance of the L1-weighted Dice Focal Loss function in comparison to standard loss functions such as Dice Loss, Dice Cross Entropy Loss and the Dice Focal Loss functions in quantifying six clinical metrics for prostate cancer disease prognosis, namely SUV\(_{\text{mean}}\), SUV\(_{\text{max}}\), TLA, TMTV, number of lesions (L), and lesion dissemination (Dmax). We also assess its generalizability across different 3D convolutional neural network (CNN)-based architectures: U-Net, SegResNet, and Attention U-Net.

\section{Materials and Methods}
\label{method}
\subsection{Dataset}
We obtained 380 PSMA [$^{18}$F]DCFPyL PET/CT scans from patients with biochemical recurrence of prostate cancer after curative treatment, recruited as part of an ongoing investigator-initiated clinical trial (NCT02899312). Ethical clearance for the data collection was obtained from the University of British Columbia – BC Cancer Ethics Board. The detailed inclusion and exclusion criteria are described in \cite{harsini2024prognostic}, but briefly, the cohort included patients with rising PSA levels indicative of disease recurrence \cite{harsini2024prognostic}. All patients who were unable to provide a written consent were excluded. Participants were administered an average dose of 350 MBq [$^{18}$F]DCFPyL corrected for body weight \cite{harsini2024prognostic}. Scans were acquired 120 minutes after injection on either a GE Discovery 600 or 690 PET/CT scanner (GE Healthcare, USA) for 2-4 minutes per bed position depending on participant size. A non-contrast-enhanced CT scan was also acquired for attenuation correction and anatomical localization.

PET images were reconstructed using an ordered subset expectation maximization algorithm with point-spread function modeling, and had a transaxial matrix size of 192 × 192 pixels. Each image contained approximately 1.80 $\pm$ 1.30 prostate cancer (PCa) lesions, with a total of 684 lesion across the entire dataset which were manually segmented by a nuclear medicine physician. The mean active lesion volume was 6.68 $\pm$ 10.20 ml, and the average  SUV\(_{\text{max}}\) and SUV\(_{\text{mean}}\) values were 12.65$\pm$14.46 and 4.62 $\pm$ 3.88, respectively.

\subsection{Ground Truth Annotation}
The ground truth segmentations were performed by a board-certified nuclear medicine physician from the BC Cancer Research Institute. All lesions were annotated using the PET Edge tool, a semi-automated gradient-based segmentation tool, and contours refined using the 3D Brush tool. Both tools are available in the MIM workstation (MIM software, Ohio, USA). These segmentations served as the ground truth labels for training the models.

\subsection{Preprocessing}
Following image acquisition, the CT images (expressed in Hounsfield Units) were clipped to a range of [-1000, 3000] before being normalized to a uniform range of [0, 1], and PSMA-PET activity concentration values in Bq/ml were converted to SUV. The PET images had an original voxel size of \(3.64 \, \text{mm} \times 3.64\, \text{mm} \times 3.27 \, \text{mm}\) and the CT images had a voxel size of \(0.98 \, \text{mm} \times 0.98 \, \text{mm} \times 3.27  \, \text{mm}\). The PET and CT images were resampled to an isotropic voxel spacing of \(2 \, \text{mm} \times 2 \, \text{mm} \times 2 \, \text{mm}\) using bilinear interpolation, and the ground truth mask was resampled to same voxel spacing using nearest-neighbor interpolation to preserve label integrity.

\subsection{Augmentation}
Prior to performing augmentation, we cropped the region outside the body in the CT, PET and GT images using a 3D bounding box. We extracted cubic patches of size \(128 \times 128 \times 128\) voxels, with an $80\%$ probability of being centered around a foreground voxel. Afterwards, we applied a series of randomized transforms exclusively to the training set which included translations within a range of (-10, 10) voxels in all spatial dimensions, rotations up to \(\frac{\pi}{15}\), and scaling up to a factor of 1.1. The augmented CT and PET patches were concatenated along the channel dimension and used as input for the segmentation network.

\subsection{Model Architecture}
Three deep neural networks with a 3D-CNN backbone, namely U-Net, Attention U-Net and SegResNet were implemented in this work. All three networks have been used in biomedical image analysis and won various competitions on segmentation of structures in medical images \cite{myronenko2022automated}\cite{cciccek20163d}\cite{oktay2018attention}. For all models, the PET and CT images were channel-concatenated and fed as input and two channel outputs were generated corresponding to lesion and background segmentation. Details of the model implementation is provided in the appendix \ref{ap:models}

\subsection{Model Training}
The dataset was split into training (323 cases) and test (57 cases) sets. The training set was further split into training and validation subsets for a 5-fold cross-validation. The model was trained to minimize the loss functions on the training set using the AdamW optimizer \cite{loshchilov2017decoupled}, with a weight decay of \(10^{-5}\). A cosine annealing scheduler was employed to gradually reduce the learning rate from \(2 \times 10^{-4}\) to zero over 1000 epochs. The networks and the training loops were implemented using the MONAI framework \cite{cardoso2022monaiopensourceframeworkdeep} and PyTorch, and the experiments were conducted on a 224 GiB RAM Tesla V-100 16GB GPU with Ubuntu 18.04, 12 CPU cores. The overall loss for an epoch, during training, was computed as the mean of the losses for across the batches. The model achieving the highest DSC on each fold was selected for use on the test set. The code is publicly available at: https://github.com/ObedDzik/pca\_segment.git.

\subsection{Loss Functions}
We trained each network with four loss functions: Dice loss, Dice Focal loss, Dice Cross Entropy and L1-weighted Dice Focal loss. The mathematical formulation of each function is detailed in appendix \ref{ap:losses}. However, we provide a description of the L1-weighted Dice Focal Loss function in this section.

\subsubsection{L1-weighted Dice Focal loss (L1DFL):}
L1DFL is a novel hybrid loss function proposed by our team that dynamically combines the Dice loss and the Focal loss. It employs an adaptive weighting mechanism, based on L1 norms (\(\Delta\)) between predicted probabilities and ground truth labels, applied to the Dice Loss term to dynamically emphasize challenging regions during training. The full details of our method is presented in \cite{dzikunu2025adaptivevoxelweightedlossusing}, and we provide a brief description here.

\begin{figure}[!t]
   \begin{centering}
   \includegraphics[width=8cm]{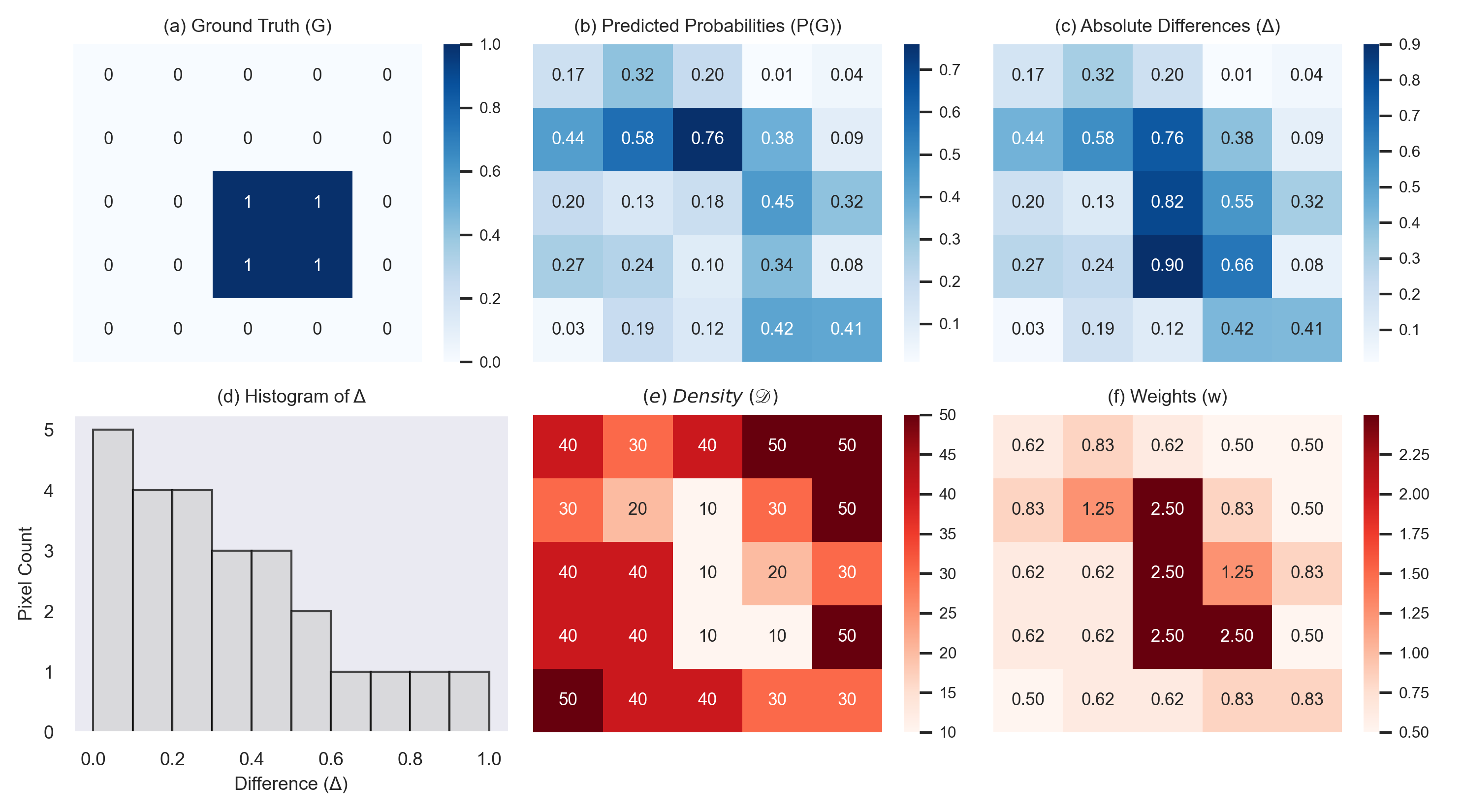}
   \caption{An illustration of the weighting strategy of the L1-weighted Dice Focal Loss function. (a) Ground truth where 0 represents background and 1 represents foreground. (b) Predicted probability map (c) L1 norms between the predicted probabilities and the ground truth labels. Values closer to 0 highlight accurate model performance and values closer to 1 highlight difficulty in pixel classification (d) shows histograms of pixel count for L1 norm values, with each bin having a width of 0.1. (e) Computed norm density based on the pixel count and bin width (f) Calculated weight based on the total number of elements and the norm density. Higher weights are assigned to regions of higher norm values and lower class frequency (mainly foreground region). The background regions and areas with lower norm values are assigned lower weights.  
   \label{fig:illustrate} 
    } 
    \end{centering}
\end{figure}

L1DFL leverages a histogram-based approach (Figure \ref{fig:illustrate}) where the computed L1 norms between predicted probabilities and ground truth labels are binned with a width of (\(\kappa\)). To account for any variations in bin sizes, particularly at the boundaries of the L1 norm value range [0,1], a normalization process is applied using an effective bin width denoted as \(\lambda(B)\), where B represents a specific bin. This normalization ensures the count is adjusted to calculate the norm density (\(\mathcal{D}(B)\)):

\begin{equation}
\mathcal{D}(B) = \frac{1}{\lambda(B)} \sum_{k=1}^n \delta_\kappa(B_k, \Delta)
\end{equation}
where the indicator function \(\delta_\kappa(\Delta_k, \Delta)\) is defined as:

\begin{equation}
\delta_\kappa(B_k, \Delta) = 
\begin{cases} 
1, & \text{if } |B_k - \Delta| \leq \frac{\kappa}{2} \\ 
0, & \text{otherwise}.
\end{cases}
\end{equation}
where \(B_k\) denotes the center of the \(k\)-th bin along the range of L1 norm values. Normalizing the computed density-norms with the total number of elements N, yields the weight \(w\):

\begin{equation}
w = \frac{N}{\mathcal{D}(B)}.
\end{equation}
The final loss function is given as: 

\begin{equation}
\text{L1DFL} = w \cdot \mathcal{L}_{\text{sDice}} + \mathcal{L}_{\text{Focal}}
\end{equation}
where $\mathcal{L}_{\text{sDice}}$ is the Dice Loss function with squared denominators \cite{milletari2016v}.

\subsection{Model Inference and Post-Processing}

On the validation and test sets, predictions were generated using the two-channel (PET and CT) whole-body images employing a sliding-window technique \cite{cardoso2022monaiopensourceframeworkdeep} with a window size of \(128 \times 128 \times 128\) voxels. This approach allowed for processing large volumetric images in smaller, manageable patches while preserving spatial information. The output predictions were resampled to align with the original ground truth (GT) mask coordinates, enabling direct comparison for the computation of evaluation metrics.

The best-performing models from each fold, based on their DSCs on the validation set, were used to predict the test set. A majority voting strategy was applied at the voxel level, where a voxel was classified as a lesion if for a network, at least three out of the five fold models predicted it as such.

\subsection{Model Evaluation}
\subsubsection{Segmentation and Detection Metrics}
\begin{figure}[!t]
   \begin{centering}
   \includegraphics[width=8cm]{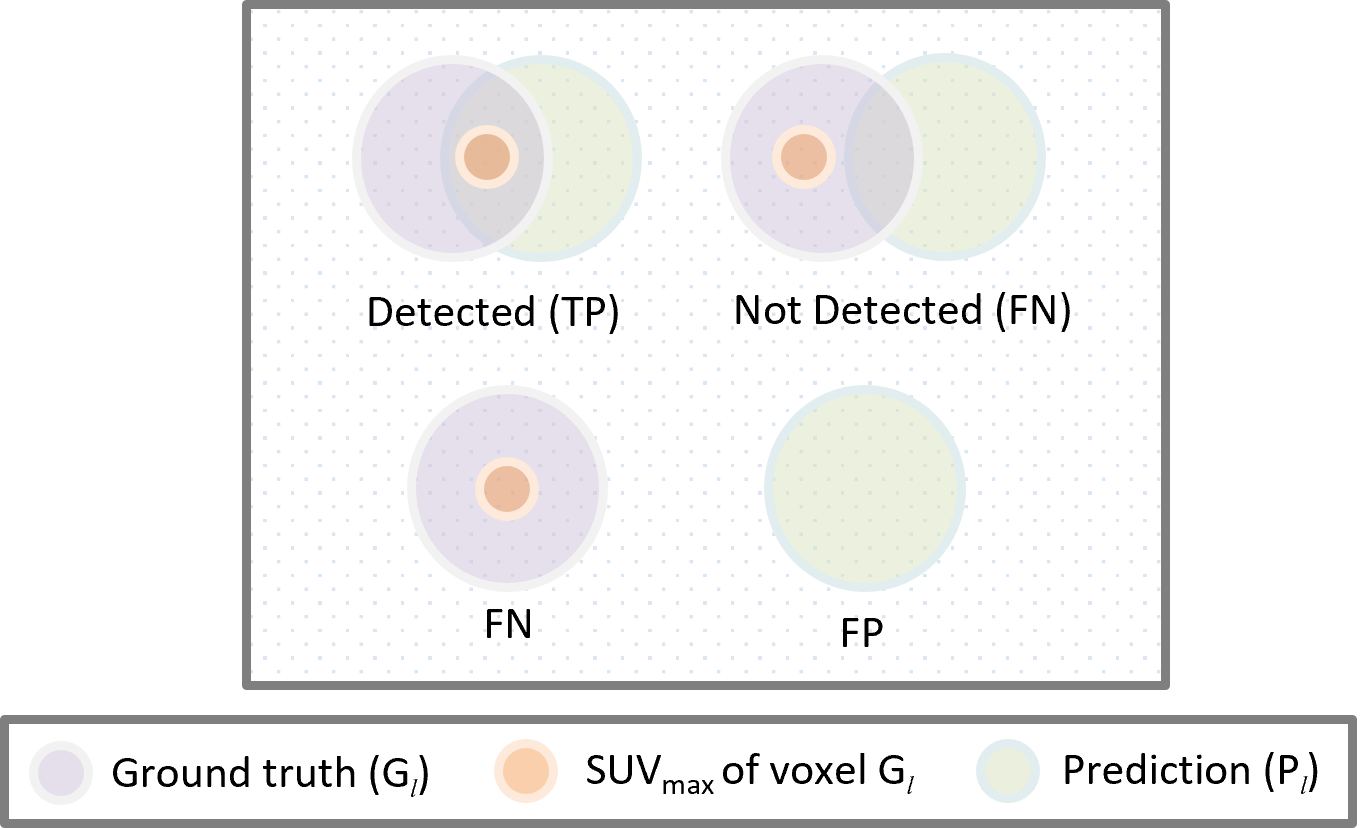}
   \caption{Illustration for defining a true positive detection based on an overlap with the voxel containing the maximum standardized uptake value (SUV\(_{\text{max}}\)) in the ground truth lesion. For a false negative (FN), either the overlap does not include the SUV\(_{\text{max}}\) voxel, or so a for a given \( G_l \) there is not a corresponding matched \( P_l \). Similarly, a false positive (FP), is a prediction, \( P_l \), for which there is no \( G_l \).  G is the set of ground truth lesions and P is the set of predicted lesions.  
   \label{fig:scheme} 
    } 
    \end{centering}
\end{figure}

We evaluate the segmentation performance of the different loss functions and the architectures based on their DSC values on the held out test set. For detection metrics, to maintain clinical relevance, we defined a true positive (TP) detection based on the inclusion of the voxel containing the maximum standardized uptake value in the overlap between a matched pair of the ground truth lesion \( G_l \) and the predicted lesion \( P_l \). By contrast, for any \( G_l \) for which there is no detection, it is designated as a false negative (FN) and a false positive (FP) was defined as a prediction \( P_l \) for which is there is no corresponding matched \( G_l \)\cite{ahamed2023comprehensive}. This definition enables the assessment of the models' performance based on their ability to detect regions of high tumor activity. We illustrate the definitions of these terms in Figure \ref{fig:scheme}. The F1 score is thus calculated as:

\begin{equation}
\text{F1} = \frac{\text{TP}}{\text{TP} + \frac{1}{2} (\text{FP} + \text{FN})}
\end{equation}

\subsubsection{Quantitative Metrics}

We assessed the performance of the various loss functions and networks via six quantitative metrics invoked in clinical analysis to probe disease progression. The mean standardized uptake value (SUV\(_{\text{mean}}\)) and maximum standardized uptake value (SUV\(_{\text{max}}\)) indicate tumor activity and aid in risk stratification and prognosis for metastatic castration-resistant prostate cancer \cite{komek2018prognostic, buteau2022psma}. Total molecular tumor volume (TMTV) is predictive of overall survival \cite{guzel2023role}, while total lesion activity (TLA), representing total tumor burden, is linked to patient outcomes and disease progression \cite{bela2024quantitative}. Lesion dissemination (Dmax), measuring lesion spread, may inform treatment decisions, and the number and distribution of lesions correlate with biochemical progression \cite{harsini2023outcome}.

\subsection{Comprehensive Evaluation of Quantitative Metrics using Statistical Methods}

Following extraction of quantitative metrics, we comprehensively evaluated agreements with ground truth using different statistical methods. To this end, we measured Lin's Concordance Correlation Coefficient (CCC), and performed Equivalence Testing as well as Bland-Altman Analysis, with respect to ground truth values for each predicted quantitative metric. In addition, we evaluated the proportion of predictions that fall within clinical margins via Coverage Probability (CP). Finally, we evaluated the maximum deviation of the predicted values from the ground truth reference using the Total Deviation Index (TDI). Thus, with CP and TDI, we assess the consistency of the models in yielding clinically relevant results while with the former three tests, we evaluate the agreement of predictions with the ground truth. For each quantitative metric, we performed a paired evaluation at the patient level, comparing the predicted value for each patient with its corresponding ground truth value in the test set. The details for each method are provided below:

\subsubsection{Concordance Correlation Coefficient (CCC)}
CCC was used to measure the level of agreement between predicted values (\(\hat{y}_i\)) and ground truth values (\(y_i\)) for each quantitativve metric, evaluating both precision and accuracy \cite{lawrence1989concordance}. The CCC is defined as:
\begin{equation}
\rho_{ccc} = \frac{2\rho \sigma_{\hat{y}_i} \sigma_y}{\sigma_{\hat{y}_i}^2 + \sigma_y^2 + (\mu_{\hat{y}_i} - \mu_y)^2},
\end{equation}
where \(\rho\) is the Pearson correlation coefficient between \({\hat{y}_i}\) and \(y\), \(\sigma_{\hat{y}_i}^2\) and \(\sigma_y^2\) are the variances, and \(\mu_{\hat{y}_i}\) and \(\mu_y\) are the means of \({\hat{y}_i}\) and \(y\), respectively. CCC takes on values in the range [-1, 1] where a value of 1 implies perfect agreement and a value of -1 implies perfect disagreement. For values between 0 and 1, we employ the interpretation suggested by \cite{mcbride2005proposal} where $CCC > 0.99$ is interpreted as an almost perfect agreement, CCC values between 0.95 and 0.99 are considered a substantial agreement, values between 0.90 and 0.95 are considered moderate agreement and values less than 0.90 are interpreted as poor.

\subsubsection{Equivalence Testing}
We conducted equivalence testing using the two one-sided t-tests (TOST) procedure with a predefined $\pm20\%$ margin, determined in consultation with physicians. Equivalence bounds were computed as \(\Delta = 0.2 \times \bar{y}\), and the null hypothesis of non-equivalence was rejected if both one-sided p-values were below \(\alpha = 0.05\). Additionally, equivalence was confirmed only if the $90\%$ confidence interval for \(\Delta\mu\) fell entirely within the predefined bounds.

\subsubsection{Bland-Altman Analysis}
We used a modified version of the Bland-Altman plots to visualize and quantify the agreement between predictions and ground truth. The values on the y-axis are the computed mean difference between the predictions and the ground truth of a given clinical metric, and the x-axis are the ground truth values of the metric. Points outside the limits of agreement indicate poor agreement and values above and below a mean difference of zero indicate instances of over and under-prediction, respectively. 

\subsubsection{Coverage Probability (CP)}
The coverage probability assesses the proportion of instances where the confidence interval encompasses the true value in a repeated run \cite{lin2002statistical}. Adopting the clinical equivalence region of \(\pm20\%\),  CP thus could be defined as the proportion of instances where the difference in predicted and ground truth values lie within the clinical range. This provides a complementary assessment to TOST where the consistency of the model's performance at the patient level is evaluated. We compute CP as:
\begin{equation}
\text{CP} = \frac{1}{N} \sum_{i=1}^N \mathbb{I}\left( \left| \hat{y}_i - y_i \right| \leq \delta \cdot y_i \right)
\end{equation}
where \(N\) is the total number of patients in the test set, \(\mathbb{I}\) is the indicator function, \(\delta\) is the acceptable clinical margin (0.2). Higher CP values reflect better precision and clinical agreement between predictions and the ground truth.

\subsubsection{Total Deviation Index (TDI)}
The Total Deviation Index (TDI) was used to quantify the deviation of predictions from the ground truth at a specified percentile (\(\tau\)) \cite{lin2002statistical}\cite{lin2000total}. This method reverses the Coverage Probability calculation, such that, given a proportion of instances (\(\tau\)), TDI evaluates the margin within which the differences fall. For this study, (\(\tau\)) was set as the 95th percentile in consultation with physicians. We calculate the TDI as:

\begin{equation}
\text{TDI}_\tau = Q_\tau\left( \{ |\hat{y}_i - y_i| \}_{i=1}^N \right)
\end{equation}
where \(Q_\tau\) is the quantile function which returns the (\(\tau\text{-th}\)) quantile of the absolute deviations, and \(\{ |\hat{y}_i - y_i| \}_{i=1}^N\) is the set of absolute deviations between the predictions \((\hat{y}_i)\) and the ground truth \((y_i)\)  values for all the N patients in the test set. Lower TDI values indicate tighter agreement between predictions and ground truth values.

\section{Results}
\label{result}

\subsection{Performance evaluation of loss functions across different networks}
\label{segmentloss}

In this section, we  provide a brief report of  segmentation and detection performances of the different networks and loss functions on the held-out test set. We provide a detailed description of the performances and the table of results in Appendix \ref{ap:evaluation}. In addition, we provide visualization of the segmentation results of some cases in Appendix \ref{ap:viz}. Overall, the L1-weighted Dice Focal Loss (L1DFL) function maintained better performance across all networks, particularly on SegResNet, where it achieved a median Dice score of $0.71$ and a mean of $0.61$, the highest among the loss functions. The percentage difference in median Dice scores between L1DFL and the Dice Cross Entropy loss function (DCE) on the SegResNet architecture is approximately $4.7\%$, with L1DFL outperforming Dice Focal by about $9.0\%$. Similar trends were observed on the Attention U-Net architecture. 

The loss functions, however, generally exhibited lower mean and median Dice scores on U-Net and Attention U-Net than on SegResNet. Despite this trend, L1DFL on the U-Net architecture, outperformed the Dice Focal Loss and Dice Loss functions by approximately $6.4\%$. However, it was marginally outperformed by the Dice Cross Entropy Loss by approximately $1.5\%$. Yet L1DFL maintained a tighter range of $0.29$ versus $0.36$ for DCE. Similarly, L1DFL had the best performance on F1 scores and false positive rates.

\subsection{Analysis of Quantitative Metrics}
\label{reproducibility}
\subsubsection{Evaluation of Concordance correlation}
\label{concordance}

\begin{figure}[!t]
   \begin{centering}
   \includegraphics[width=8cm]{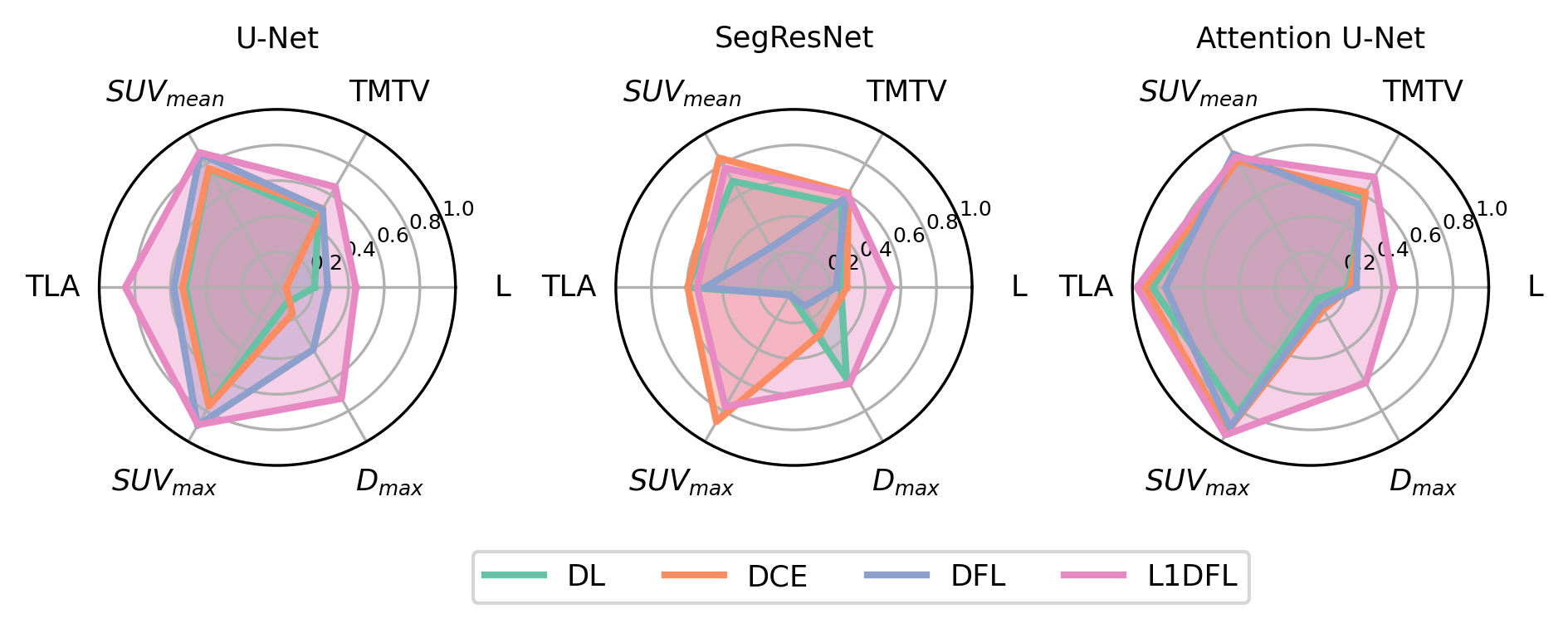}
   \caption{Radar plots showing the correlation between the predicted and ground truth metrics for the four loss functions, Dice Loss (DL), Dice Cross-Entropy (DCE), Dice Focal Loss (DFL), and L1-weighted Dice Focal Loss (L1DFL) assessed by Lin's Concordance Correlation Coefficient. The different radar plots illustrate the performance of the three architectures - U-Net, Attention U-Net, and SegResNet. The radial lines represent the correlation coefficient values. SUV\(_{\text{mean}}\): mean standardized uptake value, SUV\(_{\text{max}}\): maximum standardized uptake value, TMTV: total molecular tumor volume, TLA: total lesion activity, Dmax: lesion dissemination, L: lesion count.
   \label{fig:radar} 
    } 
    \end{centering}
\end{figure}

In this section, we assess the agreement between the predicted values of the quantitative metrics and the ground truth values using Lin's concordance correlation coefficient.  The results suggest variable performances by the different architectures and loss functions. Figure \ref{fig:radar} shows that the different loss functions achieved higher correlation coefficients on Attention U-Net and had the least values on SegResNet. Also, although not consistent across the different networks, the loss functions exhibited better agreements with the ground truth on SUV\(_{\text{max}}\), SUV\(_{\text{mean}}\) and TLA. Though overall, Dice Focal Loss exhibited the weakest correlation on SegResNet, it slightly outperformed DCE and Dice Loss on the U-Net architecture. 

L1DFL, on the other hand, obtained higher values of the correlation coefficient between 0.90-0.99. Especially on the Attention U-Net model, it outperformed others with an overall strong correlation in metrics such as TLA and SUV\(_{\text{max}}\). Meanwhile, DCE demonstrated a largely moderate performance, with weaker correlations for metrics like D\(_{\text{max}}\) and Lesion Count. However, it outperformed L1DFL, achieving a stronger correlation for SUV\(_{\text{mean}}\) and SUV\(_{\text{max}}\) with SegResNet. While the overall performance for both the Dice and Dice Focal Losses was generally poor, as most of the values of correlation fell below 0.90, showing weaker alignment with the clinical ground truth across metrics and architectures, they achieved better correlation coefficients with U-Net and Attention U-Net than with SegResNet.

\subsubsection{Test of equivalence and consistency of predictions}
\label{reproduce}

\begin{figure}[!t]
   \begin{centering}
   \includegraphics[width=8cm]{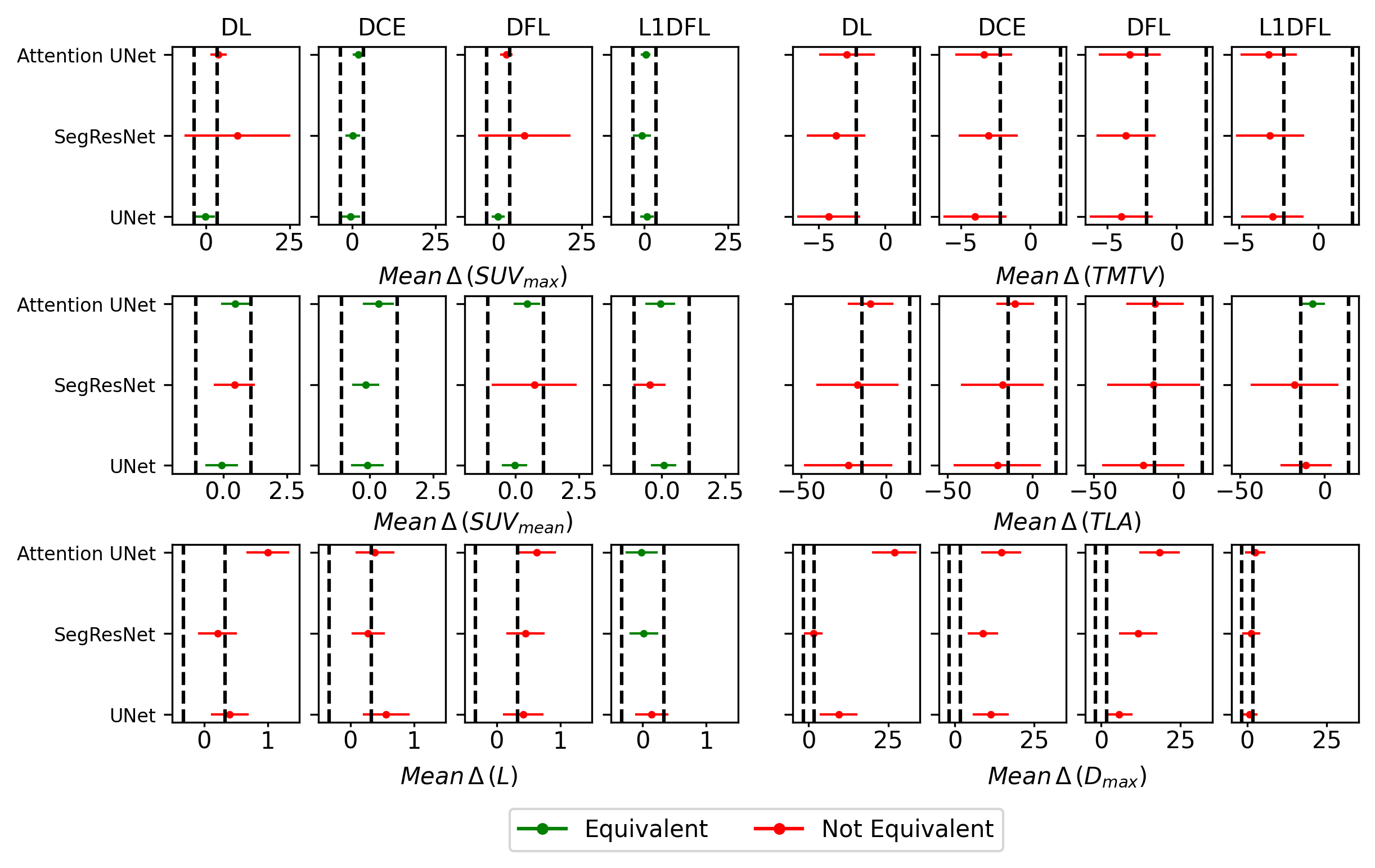}
   \caption{Two one-sided forest plots illustrating the results of equivalence testing for predicted metrics versus ground truth across different loss functions. Each row represents a specific metric, and the columns correspond to the different networks. The black vertical dashed lines represent the region of clinical equivalence ($\pm 20\% \text{ of the mean of each ground truth metric}$). The x-axis represents the mean difference between the predicted and ground truth metrics. Green bars represent equivalence between predictions of a given loss function and ground truth values at a significance level of $\alpha = 0.05$, while red bars indicate non-equivalence. 
   \label{fig:tost} 
    } 
    \end{centering}
\end{figure}

We evaluated performances of predicted quantitative measures with respect to the ground truth metrics using two one-sided t-tests (TOST), assessing equivalence at a significance level of \(\alpha = 0.05\) and an equivalence bound of \(\Delta = 20\%\). High performance was achieved for only four metrics, SUV\(_{\text{max}}\), SUV\(_{\text{mean}}\), Lesion Count (L), TLA as depicted in Figure \ref{fig:tost}, but performance was loss function and architecture dependent. In the U-Net architecture, all the loss functions were able to quantify the SUV\(_{\text{mean}}\) and SUV\(_{\text{max}}\) metrics within clinical margins; however, on SegResNet, only DCE was able to yielded acceptable quantification of both metrics. For Attention U-Net, while all loss functions reproduced SUV\(_{\text{mean}}\), only DCE and L1DFL achieved high performance for the SUV\(_{\text{max}}\) metric. Lesion Count and TLA were harder to quantify by the models, yet it was achieved by the L1DFL loss function on the Attention U-Net and SegResNet for Lesion Count and only Attention U-Net for TLA. The other metrics for which high performance was not achieved had larger confidence intervals, reflecting greater uncertainties of their estimates.

The Bland-Altman analysis across U-Net, SegResNet, and Attention U-Net architectures demonstrates that L1DFL consistently agrees better with ground truth values for all evaluated clinical metrics. For all metrics, the Dice Loss and the Dice Focal Loss functions exhibited greater variability. For instance, while L1DFL and DCE maintained tighter limits of agreement for the SUV\(_{\text{max}}\) and SUV\(_{\text{mean}}\) metrics, DL and DFL yielded larger limits mainly due to a prediction from the SegResNet architecture which had a large deviation from the ground truth value. For the other metrics, the Dice Cross Entropy loss also yielded instances of over and under prediction. That is, although DCE tended to accurately quantify the SUV\(_{\text{max}}\) and SUV\(_{\text{mean}}\) metrics, its performance on the other metrics did not always yield consistent results with the ground truth reference. In particular, the three standard loss functions often over predicted the lesion dissemination and lesion count metrics. L1DFL, on the other hand, demonstrated reduced variability and minimal bias on these metrics. This observation aligns with the TOST results where the standard loss functions failed to accurately quantify Dmax and lesion count, and only successfully predicted SUV\(_{\text{max}}\) and SUV\(_{\text{mean}}\) within acceptable margins.

For TMTV, the loss functions exhibited similar performances on the deviation from the ground truth values. With instances of over- and under-prediction across the different volume ranges, the functions struggled to maintain tight limits of agreements. This observation highlights the difficulty of models to reproduce the lesion volume since these lesions, which have resurfaced after treatment, are characterized by low uptake values and small sizes. Thus, with such small volumes, slight variations from the ground truth coupled with any instances of false predictions yield larger deviations making high fidelity performance in quantitation a challenging task.

\begin{figure}[!t]
   \begin{centering}
   \includegraphics[width=8cm]{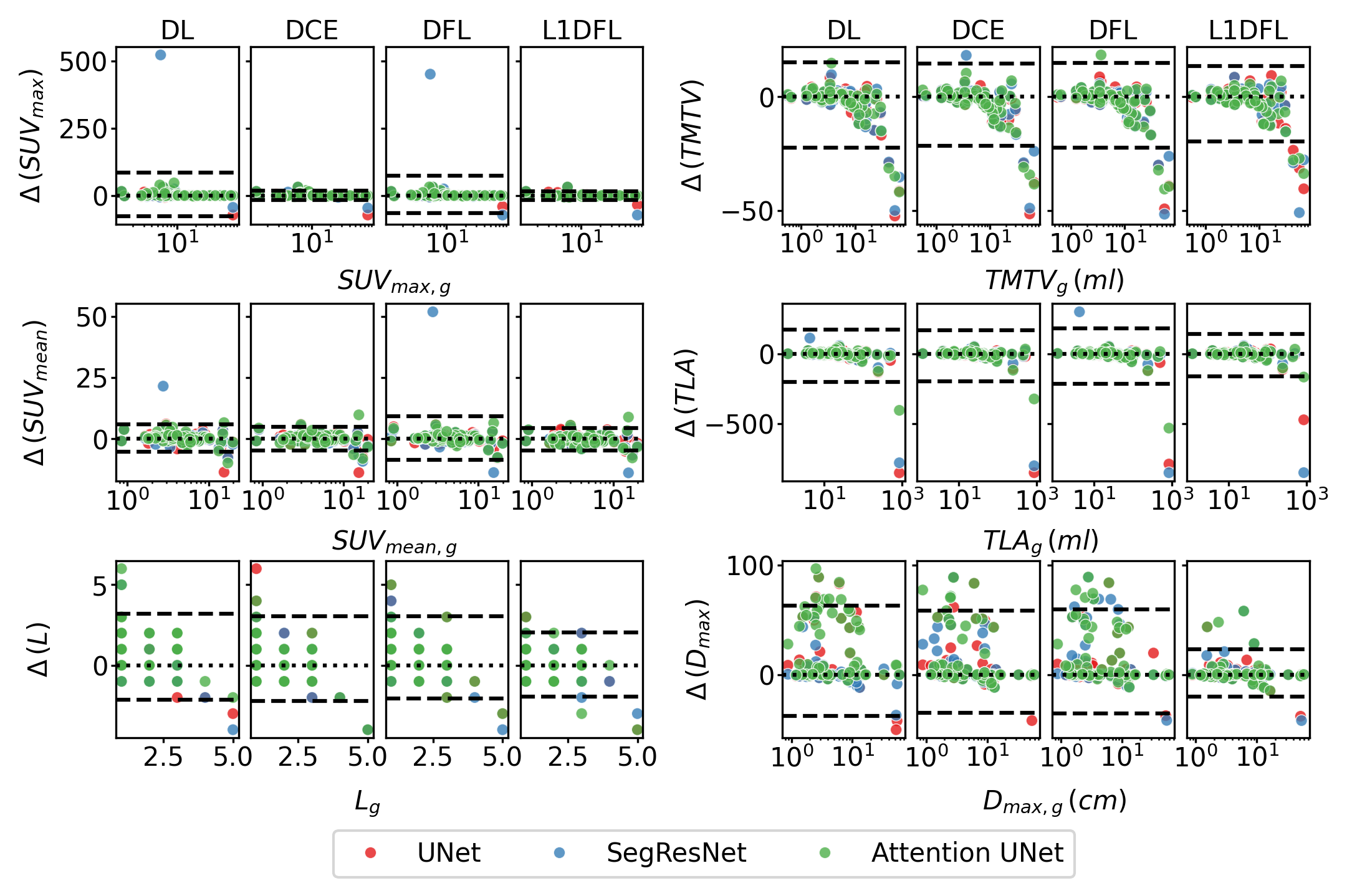}
   \caption{Figure: Modified Bland-Altman plots illustrating the variations (\( \Delta \)) between predicted and ground truth values (y-axis) against the ground truth metrics (x-axis) for different loss functions, represented by distinct colors. Each row corresponds to a specific clinical metric, and the columns represent predictions from different network architectures (U-Net, SegResNet, and Attention U-Net). The dashed black horizontal lines indicate the limits of agreement (\( \pm 20\% \) of the mean ground truth values), while the dotted line at \( \Delta = 0 \) represents perfect agreement. Each data point reflects a single instance, highlighting the distribution of variations across ground truth values.
   \label{fig:altman} 
    } 
    \end{centering}
\end{figure}

\subsection{Evaluation of performance based on coverage probability and total deviation index}
\label{cp_tdi}

We considered the metrics that were quantified by the models within acceptable clinical margins based on the two one-sided t-tests, and we evaluated the proportion of predictions that align with the ground truth metrics using coverage probability (CP) and total deviation index (TDI). In general, higher values of CP and lower values of TDI are associated with more accurate and precise predictions. 

Compared to SegResNet, U-Net, and Attention U-Net yielded generally lower CP values. However, when combined with L1DFL, Attention U-Net produced relatively higher CP values. The values obtained were all consistently higher for SUV\(_{\text{max}}\), and all the loss functions gave CP greater than 0.6, implying more than \(60\%\) of the predictions of the test set were within the clinical equivalence margin, \(\Delta =\pm 20\%\). Notably, SegResNet gave CP values over 0.8 with all loss functions for SUV\(_{\text{max}}\), showing a high portion of clinically acceptable predictions. Dice Focal Loss showed the best performance among different loss functions on U-Net, reaching a CP of 0.78 for SUV\(_{\text{max}}\), whereas L1DFL achieved this value on Attention U-Net. For lesion count, Dice Focal Loss again performed better on U-Net, reaching a CP of 0.47, equal to DCE on SegResNet. Yet both were outperformed by L1DFL with a CP of 0.54 on Attention U-Net. Lower proportions were obtained on TLA, even for L1DFL which yielded clinically acceptable quantification as assessed by the two one-side test. Thus, although the average deviation in the predicted TLA values and the ground truth fall within the clinical margin, the lower CP value suggests that the performance of L1DFL on this metric was not consistent in the different patient cases. In general, U-Net with Dice Loss had the lowest proportion of clinically equivalent predictions across the different metrics.

TDI analysis showed a similar pattern where the Dice Loss has the highest deviation on at least three metrics, SUV\(_{\text{max}}\), SUV\(_{\text{mean}}\), and Lesion count, for all architectures. The Dice Focal loss also did not yield precise predictions, with its deviation on SUV\(_{\text{max}}\) on SegResNet being almost three times the value for L1DFL. In fact, on this architecture, the Dice Focal Loss struggled to minimize inconsistent performance on all the quantitative metrics. It maintained high deviations on Attention U-Net and U-Net on SUV\(_{\text{mean}}\), L, and TLA. Contrary to the relative superior performance of DCE on the CP assessment, the Dice Focal loss' overall performance on the test set highlights instances of over and under-predictions resulting in higher deviations. It had the worse deviation on Lesion count and SUV\(_{\text{mean}}\). L1DFL on the other hand achieved the lowest deviations on all metrics, although with different architectures. The consistently higher CP values and lower TDI values for L1DFL points to its potential of maintaining predictions close to the ground truth metrics.

 \begin{figure}[!t]
   \begin{centering}
   \includegraphics[width=8cm]{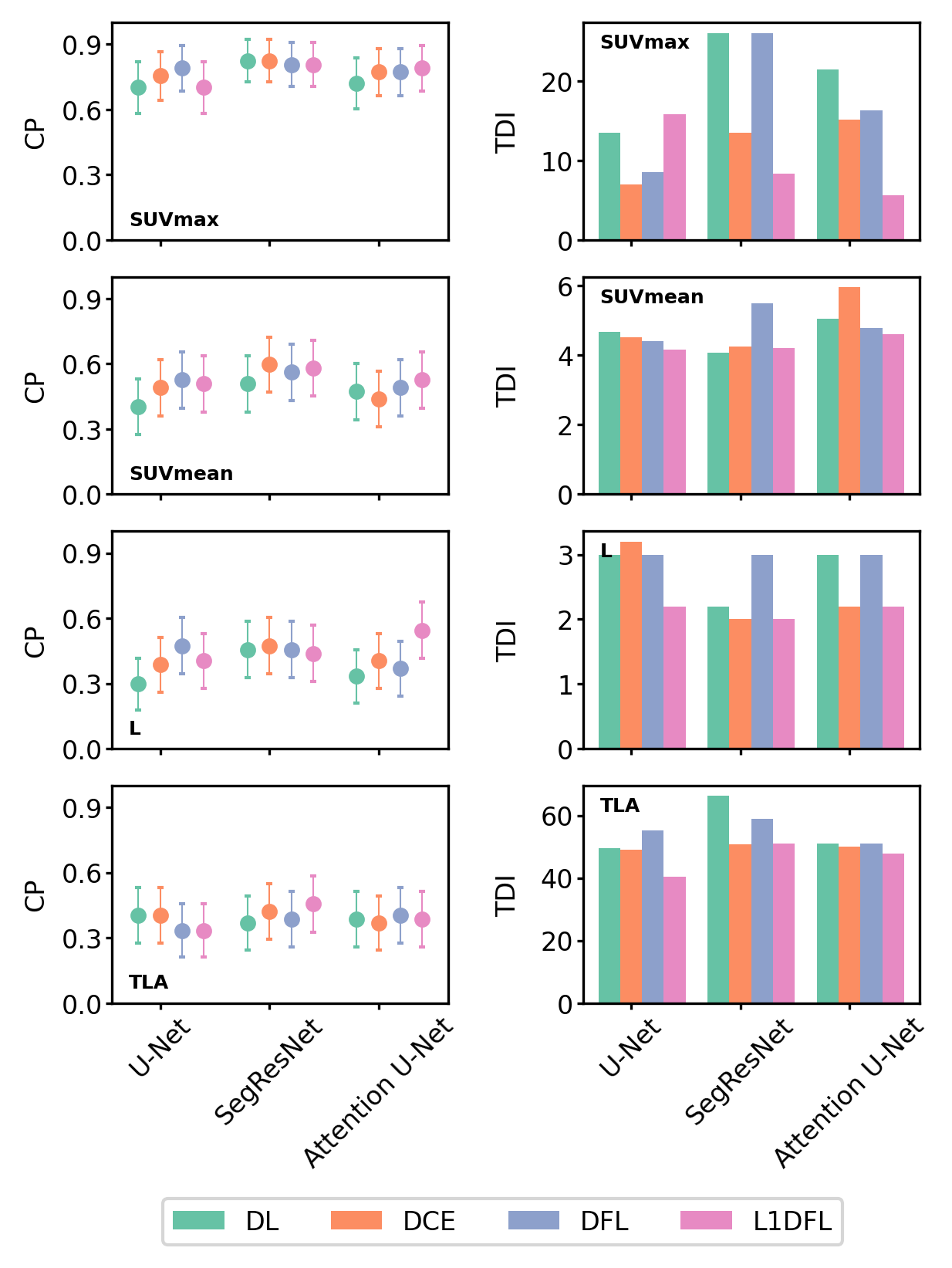}
   \caption{The left column displays plots of Coverage Probability (CP), representing the proportion of instances where the model's predictions fell within \( \pm 20\% \) of the corresponding clinical metric. Error bars indicate the $95\%$ confidence intervals. The right column shows bar plots of the Total Deviation Index (TDI) for each metric, representing the 95th percentile of deviations between predictions and ground truth values. Rows correspond to specific metrics (SUV\(_{\text{max}}\), SUV\(_{\text{mean}}\), Lesion Count (\(L\)), and TLA), while columns indicate different network architectures (U-Net, SegResNet, and Attention U-Net). Distinct colors denote different loss functions: DL, DCE, DFL, and L1DFL.
   \label{fig:curves} 
    } 
    \end{centering}
\end{figure}

\section{Discussion}
\label{discuss}

In this work, we evaluated the performance of a custom loss function, L1DFL, in predicting quantitative measures of clinical value, including SUV\(_{\text{mean}}\), SUV\(_{\text{max}}\), lesion count, Dmax, TLA, and TMTV. We compared its performance with other commonly used loss functions, Dice loss, Dice Cross Entropy loss, and Dice Focal loss on three different 3D CNN architectures, U-Net, Attention U-Net, and SegResNet. The results highlight differences in the performance of the loss functions and the various architectures based on true positive detections and their balance with false positive detection rate. The balance between both detection rates is crucial in ensuring patient care in medical applications. The results further suggest that better performance on lesion segmentation, as assessed by the Dice Similarity Coefficient, does not imply superior performance in reproducing clinical measures. For example, the loss functions achieved higher Dice scores on the SegResNet architecture than on U-Net and Attention U-Net. However, Attention U-Net had better concordance correlation coefficients on the quantitative measures. The SegResNet had the worst correlation. Correlation also varied across the different measures. Moderate to strong correlation $(CCC>0.9)$ was obtained on SUV\(_{\text{mean}}\), SUV\(_{\text{max}}\), and TLA, but a weak correlation was achieved on lesion count, Dmax, and TMTV. Poor correlation between the predictions on Dmax and lesion count and their ground truth values could be attributed to the occurrence of false positive segmentation and, in some cases, false negative detections. 

As shown in the Bland Altman plots (Figure \ref{fig:altman}), there are occurrences of overprediction on Dmax, highlighting increased distances between lesion pairs compared to the ground truth due to false positive detections. On this measure, Attention U-Net had the most increased variability in performance, with the Dice Loss and Dice Focal Loss yielding the most over prediction results. This observation is supported by their higher mean false positive rates shown on this architecture in Table \ref{tab:results}. SegResNet and U-Net also had instances of overprediction on Dmax, but they had a narrower range of variation between the predicted results and the ground truth values. Similarly, a wider deviation spread was noted in the performance on lesion count. The loss functions had instances of overprediction for a lesion count of one owing to false positive segmentations. However, there were underpredictions for a lesion count of five due to false negatives. L1DFL had the most robust performance in quantifying both the lesion count and lesion spread (Dmax), generally yielding more consistent results, followed by the Dice Cross-Entropy loss function. 

All the loss functions, however, had instances of underprediction on higher values of TMTV suggesting an increase in false negative rates. This result is supported by the observation of underestimation of lesion count, especially for a lesion count of five. Thus, the loss functions generally struggle in segmenting all lesions as the number of lesions present increases. This is expected as lesions may occupy different anatomical locations with lesion metastasis and vary in size, increasing the detection complexity. With weaker correlations obtained for TMTV, Dmax, and lesion count, the loss functions failed to reproduce these metrics. Nevertheless, L1DFL yielded acceptable quantification of lesion count on Attention U-Net and SegResNet. 

For SUV\(_{\text{mean}}\) and SUV\(_{\text{max}}\), the predictions from the four loss functions were statistically equivalent to the ground truth values, although this performance was architecture-dependent. For instance, for SUV\(_{\text{mean}}\), only L1DFL and DCE yielded equivalence to the ground truth on Attention U-Net. Similarly, only DCE reproduced SUV\(_{\text{max}}\) on SegResNet. Meanwhile, the proportion of predictions that fall within the clinically acceptable margin of $\Delta=\pm20\%$ varies even among the metrics for which acceptable quantification was achieved. Specifically, up to $90\%$ of the predictions of SUV\(_{\text{max}}\) were within the acceptable clinical margin, but this value decreased to between $30\%$ and $80\%$ for SUV\(_{\text{mean}}\) and less than $60\%$ on lesion count. Overall, L1DFL had a higher proportion of predictions in the clinical margin followed by DCE, with the least being the Dice Loss function. Besides, the Dice Loss and the Dice Focal Loss yielded higher deviations from the ground truth metrics, suggesting lower overall precision between predictions obtained on the test set and their actual values.

This study has a few key limitations. First, the dataset used in this work included a maximum of five lesions per image, limiting our ability to assess the performance of the loss functions in cases with a higher lesion burden. Therefore, the behavior of the proposed L1DFL and other loss functions in the presence of a greater number of lesions remains unexplored. Secondly, although trends are observed, larger datasets are necessary to validate the findings and achieve stronger statistical power in future studies. Additionally, the study focused on multimodal PET/CT imaging from prostate cancer patients, which may limit generalizability to other cancer types or imaging modalities. 

\section{Conclusion}
The L1-weighted Dice Focal Loss (L1DFL) function is an adaptive loss function that modifies the focus of models based on the classification difficulty of voxels, as measured by the L1 norms of predicted probabilities and ground truth labels. We utilized a medically relevant evaluation approach to assess the performance of the L1DFL function in segmenting lesions and quantifying clinical metrics that are crucial for evaluating disease progression in metastatic prostate cancer. Our analysis compared the performance of this custom loss function against that of standard Dice Loss, Dice Focal Loss, and Dice Cross Entropy Loss. We also tested its generalizability across three different architectures with CNN backbones: Attention U-Net, U-Net, and SegResNet. Among the six metrics analyzed, conventional loss functions were able to accurately quantify only the mean and maximum standardized uptake values (SUV\(_{\text{mean}}\) and SUV\(_{\text{max}}\)). By contrast, L1DFL accurately predicted the total lesion activity (TLA) and lesion count measures, alongside the standardized uptake value measures. Furthermore, L1DFL demonstrated superior performance with higher median Dice Similarity Coefficients (DSCs) and F1 scores, effectively reducing false positive detection rates.

\section{Acknowledgments}
All authors declare that they have no known conflicts of interest in terms of competing financial interests or personal relations that could have an influence or are relevant to the work reported in this paper. This work was supported by the Canadian Institutes of Health Research (CIHR) Project Grants PJT-162216 and PJT-173231. We also acknowledge computational resources and services provided by Microsoft AI for Health.

\bibliographystyle{IEEEtran}
\bibliography{reference}

\appendices
\section{Model Architecture}
\label{ap:models}

\paragraph{U-Net} The U-Net \cite{cciccek20163d} model comprised a symmetric encoder-decoder path. The encoder pathway embedded the PET/CT input volume into a lower-dimensional feature space while progressively downsampling in the spatial dimensions. Each encoder block consisted two residual units, each made up of a 3D convolutional layer, batch normalization, and ReLU activation. It achieved downsampling through convolutional layers with \(2 \times 2 \times 2\) strides, effectively reducing the spatial dimensions by half at each level, increasing feature channels from 16 to 512. The decoder path followed the modular structure of the encoder, each block comprising two residual units. The feature maps were progressively upsampled using transposed convolutions with the same stride as in the encoder path.

\paragraph{Attention U-Net} The Attention U-Net \cite{oktay2018attention} incorporates attention gates along the encoder-decoder path of the standard U-Net to provide additional focus. The network consisted five layers, where the number of channels gradually increased from 16 to 256 across the layers, and downsampling was carried out in each stage with a stride of 2. Each convolutional layer was followed by ReLU activation and Batch Normalization.

\paragraph{SegResNet} The SegResNet \cite{myronenko2022automated} architecture incorporates residual blocks with skip connections, facilitating feature propagation and ensuring stable gradients during training. We adopted an encoder consisting of blocks with 1, 2, 2, and 4 layers and for each downsampling step, the initial filter size of 16 channels was doubled. A ReLu activation and Group Normalization was applied to each convolutional layer. Each block in the decoder consisted of a single layer, reconstructing the segmentation output, while using skip connections to preserve spatial information from the encoder.

\section{Definition of Loss Functions}
\label{ap:losses}

We compared the performance of the L1-weighted Dice Focal Loss function against three other standard loss functions commonly used in the domain of medical image segmentation. We provide the mathematical description of the standard loss functions below.

\paragraph{Dice Loss (DL)}
The batch-wise loss for a cubic patch based on the Dice loss \cite{10.1007/978-3-319-46976-8_19} was computed as: 
\begin{equation}
\label{eq:DiceLoss}
   \mathcal{L}_\text{Dice} = 1 - \frac{1}{2}\sum_{c \in \{0, 1\}} \frac{2 \sum_i p_i (c) g_i(c)}{\sum_i p_i(c) + \sum_i g_i (c) + \epsilon}
\end{equation}
where  \( c \) is the class, either lesion or non-lesion, and \( p_i \) and \( g_i \) are the predicted probability and ground truth label, respectively, of a voxel \( i \). To avoid division by zero, a small constant, \( \epsilon \), is added to the denominator term. The Dice loss maximizes the overlap between the predicted segmentation and the ground truth.

\paragraph{Dice Cross-Entropy (DCE)}
The Dice Cross-Entropy loss combines the Dice loss and Cross-Entropy to form a compound loss. The loss is defined as:

\begin{equation}
\label{eq:DCE}
\mathcal{L}_{\text{DCE}} = \mathcal{L}_{\text{Dice}} + \mathcal{L}_{\text{CE}}
\end{equation}
where \( \mathcal{L}_{\text{Dice}} \) is the Dice loss, \( \mathcal{L}_{\text{CE}} \) is the Cross-Entropy loss. The Cross-Entropy loss is defined as:

\begin{equation}
\mathcal{L}_{\text{CE}} = -\frac{1}{2} \sum_{c \in \{0, 1\}}\sum_i \left[ g_i(c) \log(p_i(c))\right].
\end{equation}
where \( p_i \) is the predicted probability for voxel \( i \), \( g_i \) is the ground truth label for voxel \( i \).

\paragraph{Dice Focal Loss (DFL)}
The Dice Focal loss \cite{zhu2019anatomynet} is a compound loss function based on the combination of the Dice loss. The loss is given as:

\begin{equation}
\label{DFL}
    \mathcal{L}_{\text{DFL}} = \mathcal{L}_{\text{Dice}} + \mathcal{L}_{\text{Focal}}
\end{equation}

The Focal Loss \cite{lin2017focal} is defined as:

\begin{equation}
\label{eq:Focalloss}
    \mathcal{L}_{\text{Focal}} = -\frac{1}{2}\sum_{c \in \{0, 1\}}\sum_i \alpha_c (1 - p_i(c))^\gamma \log(p_i(c))
\end{equation}
where \( p_i \) is the predicted probability for a class \( c \) of the $i^{th}$ voxel, \( \alpha \) is the balance factor between the lesion and non-lesion classes, and \( \gamma \) is the focusing parameter that controls the rate at which easy examples are down-weighted. In our experiments, we set \( \gamma =2 \) and  \( \alpha  = 1\) following a similar implementation in \cite{10.1007/978-3-031-09002-8_9}.

\paragraph{L1-weighted Dice Focal Loss}
In the L1-weighted Dice Focal Loss function, we implement the Dice Loss function with squared denominators as opposed to the non-squared denominator version presented in \ref{eq:DiceLoss} owing to its superior performance in \cite{10.1007/978-3-030-32226-7_10}. We compute the loss, $\mathcal{L}_{\text{sDice}}$ as:

\begin{equation}
\label{eq:dice_square}
\mathcal{L}_{\text{sDice}} = 1 - \frac{1}{2}\sum_{c \in \{0, 1\}} \frac{2 \sum_i p_i (c) g_i(c)}{\sum_i p_i(c)^2 + \sum_i g_i (c)^2 + \epsilon}
\end{equation}

 For the Focal loss component of the loss function, we empirically selected \( \gamma =2 \), and for the bins, we set a constant width of 0.1 in all our experiments.

\section{Clinical Metrics}
\label{ap:metrics}
\paragraph{SUV\(_{\text{mean}}\)}
SUV\(_{\text{mean}}\) refers to the mean standardized uptake value within a segmented lesion and is representative of the activity of the lesion. It is considered an indicator of aggressiveness and response to therapy. We calculated the patient-level SUV\(_{\text{mean}}\) of a segmented lesion as follows:
\begin{equation}
\text{SUV}_{\text{mean}} = \frac{\sum_{i} P_i \cdot M_i}{\max\left( \sum_{i} M_i, 1 \right)}
\end{equation}
where \( P_i \) is the standardized uptake value at voxel \( i \) from the 3D PET image, \( M_i \) is the binary mask value at voxel \( i \) with a value of 1 if the voxel is part of a lesion, and 0 otherwise.

\paragraph{SUV\(_{\text{max}}\)}
SUV\(_{\text{max}}\) is the maximum standardized uptake value in a segmented lesion and corresponds to the most active part of the tumor. It allows for the identification of hotspots and may help guide medical interventions. The patient-level maximum SUV (\( \text{SUV}_{\text{max}} \)) is given by:
\begin{equation}
\text{SUV}_{\text{max}} = \max \left( P_i \cdot M_i \right)
\end{equation}
where \( P_i \) is the standardized uptake value at voxel \( i \) from the 3D PET image, \( M_i \) is the binary mask value at voxel \( i \) with a value of 1 if the voxel is part of a lesion, and 0 otherwise.

\paragraph{TMTV}
TMTV is the sum of the molecular tumor volumes across all lesions in an image. It is indicative of the whole tumor burden; poor prognosis is generally associated with higher TMTV. TMTV in cubic centimeters is given by:
\begin{equation}
\text{TMTV} = \left( \frac{\prod_{d=1}^3 \text{S}_d}{1000} \right) \times \sum_i M_i
\end{equation}
where \( \prod_{d=1}^3 \text{S}_d \) is the product of the voxel spacings along the three axes (x, y, z) in millimeters (mm), giving the voxel volume in cubic millimeters and converted to cubic centimeters by the division term to obtain,\(V_{\text{voxel}}\), \( \sum_i M_i \) is the total number of lesion voxels.

\paragraph{TLA}
TLA corresponds to the sum of the product of molecular tumor volume and SUV\(_{\text{mean}}\) for all lesions and represents the total activity of the tumor burden. TLA incorporates tumor volume and activity information, providing an overall quantitative measure of the disease and therapeutic response. The total lesion activity (TLA) is given by:
\begin{equation}
\text{TLA} = \sum_{i=1}^{n} \text{SUV}_{\text{mean}, i} \times V_i
\end{equation}
where \( \text{SUV}_{\text{mean}, i} \) is the mean SUV of lesion \( i \), \( V_i = V_{\text{voxel}} \times N_{\text{voxels}, i} \), \( N_{\text{voxels}, i} \) is the number of voxels in lesion \( i \) and \(V_{\text{voxel}}\) is the voxel volume in cubic centimeters.

\paragraph{Dmax}
Dmax describes the maximum distance between two lesion voxels and indicates the spatial extent of the tumor spread. Its value may affect staging, prognosis, and therapy choice, whether systemic or localized. Considering the voxel spacing, it is calculated as the maximum Euclidean distance between any two voxels in the lesion masks. Mathematically, this was expressed as:
\begin{equation}
\text{Dmax} = \frac{\max_{i,j} \| (\vec{v}_{i} - \vec{v}_{j}) \cdot \vec{s} \|}{10},
\end{equation}
where \( \vec{v}_{i} \) and \( \vec{v}_{j} \) denote the coordinates of voxels \( i \) and \( j \), \( \vec{s} \) represents the voxel spacing in millimeters, and the factor of 10 converts millimeters to centimeters.

\paragraph{Lesion Count}
The number of lesions was determined using connected component analysis on the binary lesion mask. Each connected region in the mask was identified as a separate lesion, and the total number of connected components was reported as the lesion count.

\section{Segmentation performance}
\subsection{Network and Loss function performance of lesion segmentation}
\label{ap:evaluation}

\begin{table*}[!t]
\centering{
\begin{tabular}{@{}lcccccc@{}}
\hline
Model            & {Mean Dice Score} & Median Dice Score & Mean TP & Mean FP & Mean FN & Mean F1 Score \\ \hline
\multicolumn{7}{c}{U-Net}                                                                                        \\
DL         & 0.54 $\pm$ 0.29 & 0.62 [0.36, 0.78] & 0.76 $\pm$ 0.35 & 0.73 $\pm$ 0.89 & 0.24 $\pm$ 0.35 & 0.63 $\pm$ 0.31 \\
DCE           & \textbf{0.57 $\pm$ 0.29} & \textbf{0.67 [0.44, 0.80]} & \textbf{0.77 $\pm$ 0.36} & 0.87 $\pm$ 1.24 & \textbf{0.23 $\pm$ 0.36} & 0.63 $\pm$ 0.33 \\
DFL        & 0.55 $\pm$ 0.29 & 0.62 [0.38, 0.79] & 0.75 $\pm$ 0.38 & 0.70 $\pm$ 1.02 & 0.25 $\pm$ 0.38 & 0.63 $\pm$ 0.35 \\
L1DFL             & 0.56 $\pm$ 0.29 & 0.66 [0.48, 0.77] & 0.73 $\pm$ 0.39 & \textbf{0.47 $\pm$ 0.68} & 0.27 $\pm$ 0.39 & \textbf{0.65 $\pm$ 0.34}  \\ \hline
\multicolumn{7}{c}{SegResNet}                                                                                    \\
DL         & \textbf{0.61 $\pm$ 0.25} & 0.68 [0.49, 0.80] & 0.77 $\pm$ 0.36 & 0.54 $\pm$ 1.00 & 0.23 $\pm$ 0.36 & 0.70 $\pm$ 0.34  \\
DCE           & 0.60 $\pm$ 0.27 & 0.68 [0.49, 0.80] & \textbf{0.80 $\pm$ 0.33} & 0.56 $\pm$ 0.79 & \textbf{0.20 $\pm$ 0.33} & 0.71 $\pm$ 0.30  \\
DFL        & 0.60 $\pm$ 0.26 & 0.65 [0.51, 0.79] & 0.79 $\pm$ 0.34 & 0.70 $\pm$ 0.98 & 0.21 $\pm$ 0.34 & 0.68 $\pm$ 0.33  \\
L1DFL             & \textbf{0.61 $\pm$ 0.28} & \textbf{0.71 [0.52, 0.81]} & 0.79 $\pm$ 0.36 & \textbf{0.31 $\pm$ 0.48} & 0.21 $\pm$ 0.36 & \textbf{0.72 $\pm$ 0.33}   \\ \hline
\multicolumn{7}{c}{Attention U-Net}                                                                              \\
DL         & 0.59 $\pm$ 0.26 & 0.66 [0.47, 0.80] & \textbf{0.82 $\pm$ 0.34} & 1.09 $\pm$ 1.30 & \textbf{0.19 $\pm$ 0.34} & 0.62 $\pm$ 0.31  \\
DCE           & 0.58 $\pm$ 0.27 & 0.65 [0.44, 0.78] & 0.77 $\pm$ 0.36 & 0.69 $\pm$ 0.93 & 0.23 $\pm$ 0.36 & 0.65 $\pm$ 0.33  \\
DFL        & 0.59 $\pm$ 0.27 & 0.64 [0.50, 0.81] & 0.79 $\pm$ 0.34 & 0.84 $\pm$ 1.07 & 0.21 $\pm$ 0.34 & 0.66 $\pm$ 0.30  \\
L1DFL             & \textbf{0.60 $\pm$ 0.28} & \textbf{0.67 [0.54, 0.79]} & 0.78 $\pm$ 0.37 & \textbf{0.30 $\pm$ 0.62} & 0.22 $\pm$ 0.37 & \textbf{0.73 $\pm$ 0.35}  \\ \hline
\end{tabular}
\caption{Comparison of the loss functions based on a five-fold validation, and on the test set. We report performance based on mean and median patient-level DSC. The mean DSC, TP, FP, FN and F1 scores are presented with standard deviations and the median DSC with interquartile ranges. A detection is considered a true positive if the predicted mask overlaps with the voxel containing the SUV\(_{\text{max}}\) value in the ground truth.} 
\label{tab:results}
}
\end{table*}

Table \ref{tab:results} presents the performance of various loss functions and networks on the test set. The analysis indicates that the loss functions achieved superior Dice Similarity Coefficient (DSC) values and detection rates when applied to SegResNet, compared to Attention U-Net and U-Net. However, despite achieving competitive true positive detection rates, the standard loss functions exhibited notably high false positive rates, which negatively impacted their overall mean and median DSC values. For example, Dice Loss recorded the highest mean true positive rate of $0.82$ with the Attention U-Net, closely followed by the Dice Focal Loss function; however, both suffered from poor false positive rates. Similarly, Dice Cross Entropy Loss yielded the highest mean true positive rate for U-Net, yet its lower efficacy in minimizing false positive detections resulted in a reduced F1 score compared to the L1-weighted Dice Focal Loss, which surpassed it by at least $3\%$. The L1-weighted Dice Focal Loss effectively maintained a balance between false negative and positive rates, leading to high F1 scores across different architectural types.

Furthermore, the L1DFL demonstrated tighter interquartile ranges for the Dice Similarity Coefficient (DSC), indicating that the majority of predictions fall within a narrow band of high Dice scores. Additionally, it maintained the highest DSC at the 25th percentile across all architectures. Although the Dice Cross Entropy Loss (DCE) achieved a higher median and a 75th percentile DSC compared to L1DFL on the U-Net architecture, L1DFL exhibited a narrower range of $0.29$ compared to $0.36$ for DCE. By contrast, the Dice Loss and Dice Focal Loss showed wider interquartile ranges with lower median scores, suggesting less consistent performance across individual cases. Overall, the Dice Focal Loss recorded the lowest median DSC across the different architectures. The performance of the Dice Loss and Dice Cross Entropy Loss appeared to be architecture-dependent; for instance, on the U-Net, DCE outperformed Dice Loss by at least $8\%$, whereas on the Attention U-Net, the Dice Loss marginally surpassed the Dice Cross Entropy Loss. Nonetheless, both loss functions achieved a similar DSC of 0.68 with an interquartile range of 0.31 on SegResNet. This performance, however, remained below that of L1DFL. Specifically, on SegResNet, L1DFL reached a median DSC of 0.71 [0.52, 0.81], representing a $4.4\%$ improvement over the performance of DCE and Dice Loss.

Comparatively, L1DFL had the best segmentation and detection performance on the test set. However, the one-tailed paired t-tests comparing the loss function's performance with the other loss functions resulted in p-values greater than 0.05, indicating that there is no evidence to reject the null hypothesis of no significant difference between performances.

\subsection{Visualization of loss function performance on lesions}
\label{ap:viz}
\begin{figure*}[!t] 
    \centering
    \begin{subfigure}{0.48\textwidth} 
        \centering
        \captionsetup{font=footnotesize}
        \includegraphics[width=\linewidth]{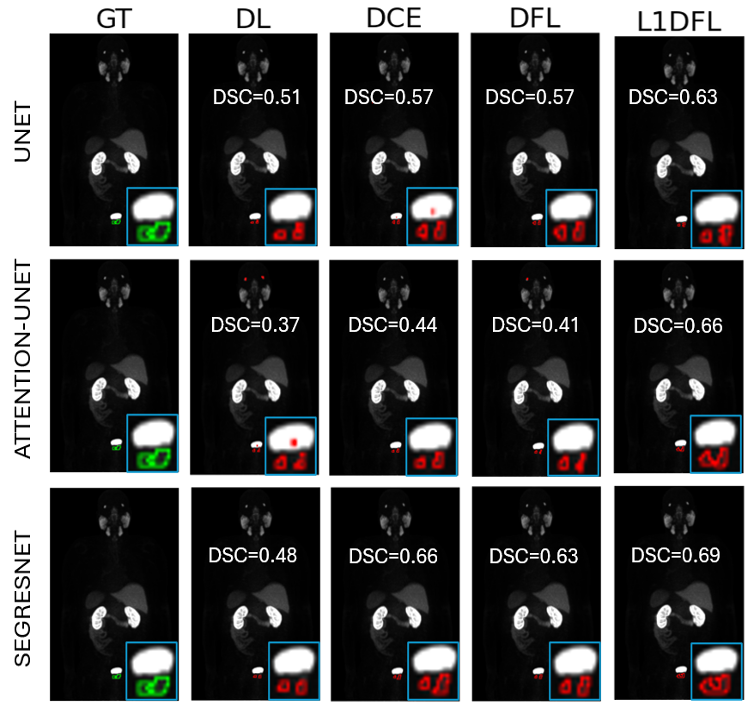}
        \caption{Segmentation performance on a lesion with a SUV\(_{\text{max}}\) of 8.12.}
        \label{ap:viz1}
    \end{subfigure}
    \hfill
    \begin{subfigure}{0.48\textwidth}
        \centering
        \captionsetup{font=footnotesize}
        \includegraphics[width=\linewidth]{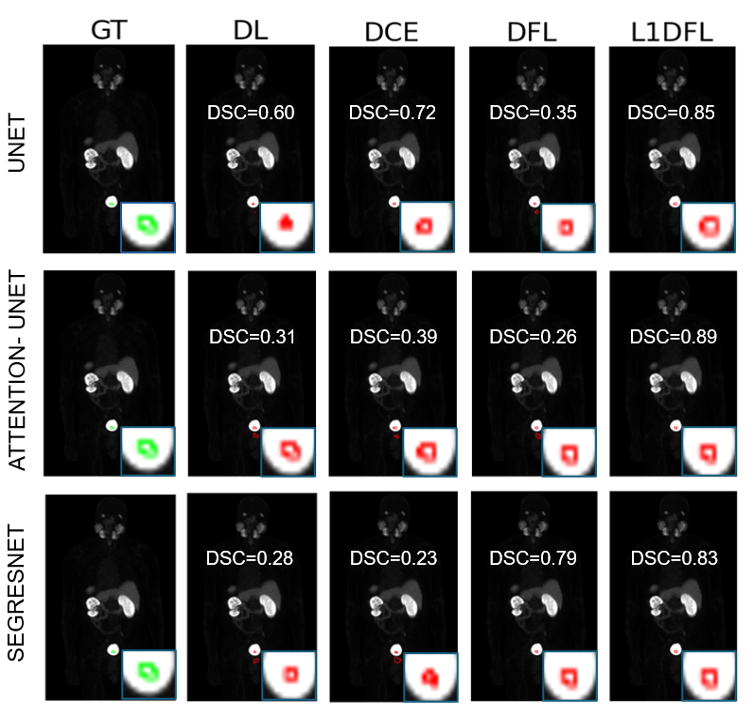}
        \caption{Segmentation performance on a lesion with a SUV\(_{\text{max}}\) of 9.66.}
        \label{ap:viz2}
    \end{subfigure}
    \caption{Comparison of segmentation performance using different loss functions on two different lesions.}
    \label{fig:comparison}
\end{figure*}

Figure \ref{ap:viz1} presents segmentation performance for a lesion with a SUV\(_{\text{max}}\) of 8.12. while \ref{ap:viz2} illustrates performance for a lesion with a SUV\(_{\text{max}}\) of 9.66. Each visualization compares the performance across different loss functions paired with each architecture based on SUV\(_{\text{max}}\) categorization.

\vfill

\end{document}